%% file: main.tex
\documentclass[review]{elsarticle}

\usepackage[american]{babel}
\usepackage[utf8]{inputenc}
\usepackage{enumerate}
\usepackage{setspace}
\usepackage{pgfgantt}
\usepackage{booktabs}
\usepackage{listings}
\usepackage{hyperref}
\usepackage{xspace}
\usepackage{todonotes}
\usepackage{mdframed}
\usepackage{dcolumn}

\usepackage{csvsimple}
\usepackage{pgfplotstable,filecontents}

\newmdtheoremenv [
 outerlinewidth = 1 ,
 roundcorner = 1pt,
 leftmargin = 1,
 rightmargin = 1,
 backgroundcolor = gray!20,
 outerlinecolor = blue!70!black,
 ntheorem = false,
] {finding}{Finding}

\DeclareFixedFont{\ttb}{T1}{txtt}{bx}{n}{12} 
\DeclareFixedFont{\ttm}{T1}{txtt}{m}{n}{12}  

\definecolor{keywords}{rgb}{0.5,0,0.35}
\definecolor{comments}{RGB}{0,0,113}
\definecolor{red}{RGB}{160,0,0}
\definecolor{green}{RGB}{0,150,0}
 
\usepackage{fullpage}      
\usepackage{indentfirst}   

\usepackage{graphicx}       
\usepackage{subfig}
\usepackage{venndiagram}

\newcommand{\joke}{Joker\xspace}
\newcommand{\blls}{BLL-Study\xspace}

\begin{document}

\begin{frontmatter}
\title{Exploring the Use of Static and Dynamic Analysis to Improve the Performance of the Mining Sandbox Approach for Android Malware Identification}

\author[1]{Francisco Handrick da Costa}
\author[1]{Ismael Medeiros}
\author[1]{Thales Menezes}
\author[1]{Jo\~{a}o Victor da Silva}
\author[1]{Ingrid Lorraine da Silva}
\author[1]{Rodrigo Bonif\'{a}cio}
\author[2]{Krishna Narasimhan}
\author[3]{M\'{a}rcio Ribeiro}

\address[1]{Computer Science Department, University of Bras\'{i}lia, Brazil}
\address[2]{Software Technology Group, TU Darmstadt, Germany}
\address[3]{Institute of Computing, Federal University of Alagoas, Brazil}

\input{abstract}

\end{frontmatter}

\input{introduction}
\input{background}
\input{setup}
\input{results}
\input{implications}

\input{threats}
\input{conclusion}

\bibliographystyle{elsarticle-num}
\bibliography{main}

\end{document}

%% file: abstract.tex
\begin{abstract}
The popularization of the Android platform and the growing number of Android applications (apps) that manage sensitive data turned the Android ecosystem into an attractive target for malicious software. For this reason, researchers and practitioners have investigated new approaches to address Android's security issues, including techniques that leverage dynamic analysis to \textit{mine Android sandboxes}. The mining sandbox approach consists in running dynamic analysis tools on a benign version of an Android app. This exploratory phase records all calls to sensitive APIs. Later, we can use this information to (a) prevent calls to other sensitive APIs (those not recorded in the exploratory phase) or (b) run the dynamic analysis tools again in a different version of the app. During this second execution of the fuzzing tools, a warning of possible malicious behavior is raised whenever the new version of the app calls a sensitive API not recorded in the exploratory phase. 

The use of a mining sandbox approach is an effective technique for Android malware analysis, as previous research works revealed. Particularly, existing reports present an accuracy of almost 70\% in the identification of malicious behavior using dynamic analysis tools to mine android sandboxes. However, although the use of dynamic analysis for mining Android sandboxes has been investigated before, little is known about the potential benefits of combining static
analysis with a mining sandbox approach for identifying malicious behavior.
Accordingly, in this paper we present the results of two studies that investigate the impact of using static analysis to complement the performance of existing dynamic analysis tools tailored for mining Android sandboxes, in the task of identifying malicious behavior.

In the first study we conduct a non-exact replication of a previous study (hereafter \blls) that compares the performance of test case generation tools for mining Android sandboxes. Differently from the original work, here we isolate the effect of an independent  static analysis component (DroidFax) they used to instrument the Android apps in their experiments. This decision was motivated by the fact that DroidFax could have influenced the efficacy of the dynamic analyses tools positively---through the execution of specific static analysis algorithms DroidFax also implements. In our second study, we carried out a new experiment to investigate the efficacy of taint analysis algorithms to complement the mining sandbox approach previously used to identify malicious behavior. To this end, we executed the FlowDroid tool to identifying source-sink paths from benign/malign pairs of Android apps used in a previous research work. 

Our study brings several findings. For instance, the first study reveals that DroidFax alone (static analysis) can detect 43.75\% of the malwares in the \blls dataset, contributing substantially in the the performance of the dynamic analysis tools in the \blls. The results of the second study show that taint analysis is also practical to complement the mining sandboxes approach, with a performance similar to that reached by dynamic analysis tools.
\end{abstract}


%% file: introduction.tex
\section{Introduction}\label{sec:introduction}

Almost two-thirds of the world use mobile technologies~\cite{Comscore}, and the Android Operating System has dominated the market of smartphones, tablets, and others electronic devices \cite{statcounter}. Due to this growing popularity, the number of incidents related to Android malicious software (malware) has significantly increased. In only three years, researchers reported a substantial increase in the population of Android malwares: from just three families and a hundred samples in 2010 to more than a hundred families with thousands of samples in 2013~\cite{DBLP:journals/comsur/FarukiBLGGCR15,DBLP:journals/csur/SufatrioTCT15}. Security issues in Android software applications~\footnote{In this paper, we will use the terms Android Applications, Android Apps and Apps interchangeably, to represent Android software applications} have become a relevant research topic, and many techniques have been developed to identify vulnerabilities in Android apps~\cite{DBLP:conf/pldi/ArztRFBBKTOM14}, including the use of static analysis algorithms either to identify privacy leaks or to reveal the misuse of cryptographic primitives~\cite{krueger:ecoop-2018,rahaman:ccs-2019}, for instance.

Another alternative for protecting users from Android malicious behavior consists in the use of dynamic analysis to mine Android sandboxes~\cite{DBLP:conf/icse/JamrozikSZ16}. The mine sandbox approach starts with an
exploratory phase, in which a practitioner takes advantage of automatic test case generator tools that explores an Android application while recording the set of sensitive APIs the app calls. 
. This set of senstivie calls comprises a sandbox infrastructure. After the exploratory phase, the sandbox might then monitor any call to sensitive APIs while a user is running the app, blocking the calls that have not been identified during the exploratory phase---thereby protecting Android users from additional malicious behavior~\cite{DBLP:conf/icse/JamrozikSZ16}.
Jamrozik et al. argue in favor of dynamic analysis for mining sandboxes, instead of using static analysis---mostly because of the overapproximation problem: ``static analysis often assume that more behaviors are possible than actually would be''~\cite{DBLP:conf/icse/JamrozikSZ16}. In addition, code that uses dynamic features (such as reflection) poses additional challenges to static analysis algorithms---even though \emph{dynamic features} of programming languages are often used to introduce malicious behavior. Even though these claims are reasonable, previous research results do not present empirical assessments about the limitations of static analysis to mine sandboxes. Consequently, it is not clear whether and how both approaches (dynamic and static analysis) could complement each other in the process of mining Android sandboxes.

The lack of understanding about static and dynamic analysis complementing each other also appears in the work of Bao et al.~\cite{DBLP:conf/wcre/BaoLL18} (hereafter \blls), which presents an empirical study that explores the performance of dynamic analysis for identifying malicious behavior using the mining sandbox approach. Their study leverages DroidFax~\cite{DBLP:conf/icsm/CaiR17a} to instrument $102$ pairs of Android apps (each pair comprising a benign and a malicious version of an App) and to collect the information needed to mine sandboxes (that is, the calls to sensitive APIs).
Although the authors report a precision of at most 70\% of dynamic analysis tools to differentiate the benign and malicious versions of the apps, the authors ignore the fact that DroidFax statically analyzes the Android apps and also records calls to sensitive APIs (besides instrumenting the apps). As we discuss in this paper, this DroidFax static analysis component leads to an overestimation of the performance of the dynamic analysis tools for mining sandboxes and might have introduced a possible threat to the conclusions of that work. In the security domain, overestimating the performance of a technique for malware identification brings serious risks, and we show here that DroidFax inflated significantly the performance of the dynamic analysis tools for mining sandboxes, as reported in the \blls.

The goal of this paper is two fold. First we present the results of an
external, non-exact replication~\cite{role-of-replication} of the \blls. To this end,
we take advantage of DroidXP, a tool suite that helps researchers (including ourselves) to
integrate test case generation tools and compare their performance on
mining Android sandboxes. We discussed the design and implementation of DroidXP in a conference
paper~\cite{DBLP:conf/scam/CostaMCMVBC20}, which also
includes an initial evaluation of DroidXP.
As a matter of fact, the results of the first DroidXP evaluation revealed a possible
overestimation in the performance of dynamic analysis tools as
reported in the \blls---which in the end motivated us to
conduct the non-exact replication of that study. Here we extend
our previous work with a couple of customizations of DroidXP, which allowed us
to reproduce the \blls by means of a serie of new experiments
that reveal the actual performance of the
dynamic analysis tools. Section~\ref{sec:droidxp} revisit the
DroidXP design, while Section~\ref{sec:set1} discuss
the setup of our replication study.

Second, in this paper we also explore how a static analysis approach
(based on taint analysis) compares and complements the mining sandbox technique
for identifying malicious behavior that infects benign applications.
The idea here is to compare the dataflows from \emph{source} to
\emph{sink} statements computed using two executions of the
FlowDroid infrastructure~\cite{arzt:pldi-2014}: one execution that analyses
a benign version of an Android app and one execution that
analyses a malicious version. We consider that
the taint analysis approach is able to identify a malware whenever
we find a dataflow from a source to a sink in the second execution
that does not appear in the first one. We detail the
settings of this taint analysis study in Section~\ref{sec:set2}

Altogether, this paper brings the following contributions:

\begin{itemize}
\item A replication of the \blls that better clarifies the performance of
  dynamic analysis tools for mining Android sandboxes. The results of
  our replication (Section~\ref{sec:res-fs})
  give evidence that the previous work overestimated
  the performance of the dynamic analysis tools---that is, without
  DroidFax (an independent component used for running the
  \blls experiment), the performance of the tools drop between $16.44$\% to $58$\%. 

\item A broad comprehension about the role of static analysis tools for mining
  sandboxes, showing that we can benefit from using both static and dynamic
  analysis for detecting malicious Android apps. In addition,
  we give evidence that a well known static analysis approach, based on
  taint analysis, leads to a performance similar to the dynamic analysis
  approach for diferenciating benign and malicious versions of the same
  app (Section~\ref{sec:res-ss}).

\item A reproduction package of our study that is available online, including
  scripts for statistic analysis \footnote{https://htmlpreview.github.io/?https://github.com/droidxp/paper-replication-package/blob/master/replication.html}
  and tooling for reproducing and extending our study. The repository for DroidXP is available
at GitHub\footnote{https://github.com/droidxp/benchmark}.
 
\end{itemize}

%% file: background.tex
\section{Background and Related Work}

In this section, we introduce the concepts and terminology that are necessary to understand the remainder of this paper. First, Section~\ref{sec:sandbox} introduces some background information about the use of \emph{sandboxes} to protect invalid access to sensitive resources. After that, in Section~\ref{sec:android-sandbox} we review the mining sandbox approach for detecting malicious behavior in Android apps. Finally, Section~\ref{sec:taint} presents some background information about taint analysis.

\subsection{The Sandbox Approach for Protecting Resources}\label{sec:sandbox}

A \emph{sandbox} is an isolated environment on an electronic device within which applications cannot affect other programs outside its boundaries, like the file system, the network, or other device data~\cite{DBLP:journals/peerj-cs/MaassSCS16}. Sandboxes enable testing and execution of unsafe or untested code, possible malware, without worrying about the integrity of the electronic device that runs the application~\cite{DBLP:conf/esorics/BordoniCS17}. This need might arise in a variety of situations, such as when executing software input by untrusted users, in malware analysis, or even as a security mechanism in case a trusted system gets compromised~\cite{DBLP:journals/peerj-cs/MaassSCS16}.
A sandbox environment must be able to shield the host machine or operating system from any damages caused by third-party software. Thus, a sandbox environment should have the minimum requirements to run programs (make sure the program will not impact resources outside the sandbox), and make sure it will never assign the program greater privileges than it should have, working with the principle of \emph{least privilege}, giving permissions to users according to their needs, i.e., giving them no more power than needed to successfully perform their task. This principle prevents escalating privileges and unauthorized access to resources, thereby improving the system's overall reliability.

Within the Android ecosystem, least privilege is realized through sandboxing process, where apps never access the data of other apps, and an app just accesses user resources, like contacts and location, through specific APIs (Application Programming Interface), which are in-turn guarded by permissions. Google Play Store is the primary market source for Android apps, and has a flexible policy regarding the apps' publishment process. Therefore, every month administrators remove several Android apps from the Play Store because of issues related to spyware and other types of malware \cite{DBLP:conf/msr/WangLL0X18}. For security reasons, Google Play lists each app with its requested permissions. However, many malicious apps usually ask for more permissions than their APIs normally would require~\cite{DBLP:conf/ccs/FeltCHSW11}. Those permissions are presented to the user during a new app's installation, since Android version 6, but most users are careless since they are only interested in the end product~\cite{DBLP:conf/soups/FeltHEHCW12}. 

Nowadays, malware becomes more stealthy and hackers learn how to avoid anti-virus signature checks, for instance by obfuscating calls to native code that is allowed to make system calls~\cite{DBLP:journals/corr/abs-2002-04540} or conducting side attacks to make system calls from a benign app.

\subsection{Mining Android Sandbox}\label{sec:android-sandbox}

The mining Android sandbox approach~\cite{DBLP:conf/icse/JamrozikSZ16} relies on test generator tools to explore an Android app's dynamic behavior, and thus mine a set of sensitive resources the app needs. The sandbox uses this set of sensitive APIs to ensure the app execution's security by restricting the resources that are allowed to use. The mining sandbox approach works in two phases. In the first, named exploratory phase, a practitioner uses test generator tools to execute a benign version of an app and record the set of sensitive APIs the app calls. In the second phase, named execution, the sandbox constraints the app to access only the sensitive APIs mined in the first phase. Accordingly, the sandbox ensures that a malicious app could not call any sensitive API, besides those calls to APIs recorded in the first phase.

The idea of automatically mining software resources or components to infer behavior is not new, and has been discussed before. For instance, Whaley et al.~\cite{DBLP:conf/issta/WhaleyML02} combine dynamic and static analysis for API mining and so infer program behavior based on an usage example of Java classes. Ammons et al.~\cite{DBLP:conf/popl/AmmonsBL02} propose a machine learning approach, called specification mining, to discover temporal and data-dependence relationships that a program follows when interacting with an API or abstract data types.

The main purpose of a test generator tool is to program crashes or bugs in general. Nonetheless, it is also possible to use test generator tools to explore program behavior (dynamic analysis), and thus assist in the task of building sandboxes. Regarding test generator tools used for mining Android sandboxes, Jamrozik et al.~\cite{DBLP:conf/icse/JamrozikZ16} proposed DroidMate, a test generation tool that implements a pseudo-random graphical user interface (GUI) exploration strategy, and was the first approach to leverage test generation to extract sandbox rules from apps. Li et al.~\cite{DBLP:conf/icse/LiYGC17} proposed DroidBot, a test generator tool that explores sensitive resources access from Android apps, following a model-based exploration strategy. In their work, the authors present a comparison between DroidBot and Monkey~\cite{Monkey} regarding malware analysis and show that DroidBot can trigger several sensitive calls more often than Monkey. Sensitive calls in the Android context occurs when an Android app functionality can result in accessing or leaking of Android users' sensitive data. Examples of sensitive calls access user location or network information. Humanoid is another test generator tool for Android~\cite{DBLP:conf/kbse/LiY0C19}---actually a DroidBot evolution. It is also a GUI test generator that learning how humans interact with Android apps. In contrary to random input generators, Humanoid uses a learned model to generate human-like test inputs, and prioritize the possible interactions on a GUI, according to their importance.

Bao et al.~\cite{DBLP:conf/wcre/BaoLL18} present a comparative study test generator tools to identify malicious behavior using the mining sandboxes approach. Their study indicates that the tools were efficient in identifying at most $70$\% of the malware in a specific dataset and also reports that after, combining all test generator tools, it was possible to detect $75.49$\% of malicious behavior explored ($77$ among $102$). However, this study did not focus on the possible interference of static analysis in the final result, since this study used DroidFax~\cite{DBLP:conf/icsm/CaiR17a} to instrument the apps, though, as we discuss in this paper, DroidFax also performs a static analysis of the apps whose results complement the dynamic analysis approach for mining sandboxes.

\subsection{Taint Analysis}\label{sec:taint}

Taint analysis is a special type of static or dynamic analysis that aims to track data flows within programs~\cite{DBLP:conf/sigsoft/PauckBW18}. Typically, taint analysis is used to identify sensitive information leakage by detecting taint flow between ``sources'' and ``sinks''. In the context of Android apps, a data leak occurs when sensitive data, such as contact, or device ID, flows from a sensitive resource to a method that might \emph{sink} information to other peers, like sending a message. Taint analysis can present possible malicious data flow to malware detection tools or even for a human check, which can decide if the ``source-sink" relationship is or is not an unwanted behavior. Thereby, taint analysis monitors sensitive sources ``tainted" through the app by starting at a pre-defined point. 

In the Android context, sources are the APIs in which apps access sensitive information, called sensitive APIs. The analysis follows the data flow until it reaches a sink, like a method that sends SMS. It brings precise information about which data will be leaked~\cite{DBLP:conf/pldi/ArztRFBBKTOM14}. The Android SDK provides APIs that allow apps to send private data to other apps on the same device, or remote devices. As these APIs may lead to sensitive data leakage, they are security-critical and require special attention and control~\cite{DBLP:conf/osdi/EnckGCCJMS10}. (Listing~\ref{lst:sourceSink}) presents a simple data leakage example. In this example, the device information is captured at line 4 (source) and then leaked at line 9 (sink), by SMS transmission.

\begin{lstlisting}[caption={Simple Data Leakage},
      language=Java, basicstyle=\fontsize{8}{6}\selectfont\ttfamily,
      label={lst:sourceSink}]

1 > localObject2 = (TelephonyManager)getSystemService("phone");
2 > if (localObject2 != null)
3 > {
4 >  this.imei = ((TelephonyManager)localObject2).getDeviceId();//source
5 > }
6 > if ("".equals(this.destMobile)) {
7 >  getDestMobile();
8 > }
9 > sendSMS(this.destMobile, "imei:" + this.imei)//sink
\end{lstlisting}

Wei et al.~\cite{DBLP:conf/issta/HuangDMD15} propose a scalable taint analysis for Android apps that applies traditional taint analysis techniques with targeted optimizations specific to Android OS. FlowDroid~\cite{DBLP:conf/pldi/ArztRFBBKTOM14} improves the precision of traditional approaches by including context and flow sensitivity. A significant issue with taint analysis is the cost of the tool itself hampering the performance. FastDroid~\cite{DBLP:journals/compsec/ZhangTD21} mitigates this issue by introducing an intermediate light-weight abstraction to perform the analysis, called taint value graph (TVG). To improve efficiency and precision, FastDroid focuses on exploring the propagation of taint values, rather than the traditional data flow analysis. FastDroid constructs taint value graphs (TVGs) exploring taint values, then it extracts a subset of potential taint flows (PTFs) from it. FastDroid improves the analysis process by performing analysis only on (PTFs). In this paper we investigate the use of taint analysis to identify malicious behavior, by mining the source an sink pairs from distinct versions of an app.

%% file: setup.tex
\section{Study Settings}

Our research work aims to better understand
the use of static analysis to mine Android sandboxes
and explore the benefits of combining taint analysis
with the mine sandbox approach for identifying malicious
behavior. 
On the one hand, Jamrozik et al. suggest that a 
static analysis approach for mining sandboxes
might be ineffective---due to \emph{overapproximation
problem}~\cite{DBLP:conf/icse/JamrozikSZ16}.
However, to the best of our knowledge,
there is no empirical study comparing
static and dynamic analysis for mining sandboxes.
On the other hand, the \blls explored the mining sandbox approach by comparing
the performance of five {\bf dynamic analysis tools} (DroidMate, DroidBot, PUMA,
GUIRipper, and Monkey) for identifying
malicious behavior. Nonetheless, their
research also involved an external static analysis component (DroidFax)
whose impact on the results was not measured---in terms of malware
identification. 
This lack of understanding about the implications of
static analysis for mining sandboxes motivates
our research, which investigates the following research questions.

\begin{enumerate}[(RQ1)]

\item What is the impact of the DroidFax static analysis algorithms on the results of the \blls?
  We estimate the impact in terms of the number of detected malwares.

 \item What is the effective performance of each sandbox, in terms of the number of detected malware, when we
   discard the contributions of the DroidFax static analysis algorithms?

 \item What are the benefits of using taint
   analysis algorithms to complement the dynamic analysis approach for mining sandboxes,
   in terms of additional malwares identified?
\end{enumerate}


Answering the research questions RQ1 and RQ2 allows us to better
understand the relevance of combining static
and dynamic analysis for mining Android sandboxes. Moreover,
exploring RQ1 and RQ2 can reveal 
a possible overestimation of the performance of the
dynamic analysis tools in the \blls. Answering research question RQ3
allows us to open up the possibility of finding new strategies for malware detection, complementing the performance of
dynamic analysis through the use of static analysis algorithms.
We conducted two empirical studies to answer the research questions above. We address
the research questions RQ1 and RQ2 in the
first empirical study, whose goal is to conduct a non-exact replication of
the \blls. We conduct the first empirical study using
DroidXP~\cite{DBLP:conf/scam/CostaMCMVBC20} (Section~\ref{sec:droidxp}), a
tool that simplifies the execution of experiments
that compare the performance of dynamic analysis tools
in the task of identifying malwares, using a mining sandbox
approach. We present the study settings of the first
empirical study in Section~\ref{sec:set1}.
In the second empirical study we use
FlowDroid~\cite{DBLP:conf/pldi/ArztRFBBKTOM14} to investigate the 
suitability of taint analysis algorithms to complement the mining sandbox
approach for identifying malwares, and thus it targets our third
research question (RQ3). We present the settings of the second
empirical study in Section~\ref{sec:set2}.

\subsection{The DroidXP benchmark}\label{sec:droidxp}

We designed and implemented DroidXP to systematically assess and compare the
performance of test generation tools for mining android sandboxes. It allows
the integration and comparison of test case generation tools for mining sandboxes, and simplifies
the reproduction of the studies. DroidXP relies on a
simple \emph{Command Line Interface} (CLI) that simplifies the integration
of different test generation tools and favors the setup and execution 
of the experiments. DroidXP also relies on DroidFax, which instruments
Android apps and collects relevant information about
their execution, including the set of sensitive APIs a given
app calls during a test execution. DroidFax also collects
inter-component communication (ICC) using  static
program analysis.

The DroidXP CLI provides commands for listing all test case
generation tools (executing the project with the option ``list-tools'') that had been
integrated into the tool and commands for executing the experiments. An
\emph{experiment run} can be configured according to several parameters, including:

\begin{itemize}
    \item \texttt{-tools}: Specifies the test tools used in the experiment
    \item \texttt{-t}: Specifies the threshold (in seconds) for the execution time in the experiment
    \item \texttt{-r}: Specifies the number of repetitions used in the experiment
    \item \texttt{-output-format}: Specifies the output format
    \item \texttt{--debug}: Specifies to run in DEBUG mode (default: false)
    \item \texttt{--disable-static-analysis}: Disable DroidFax static analysis phase (default: false)
\end{itemize}

\begin{figure*}[thb]
  \includegraphics[width=1\textwidth]{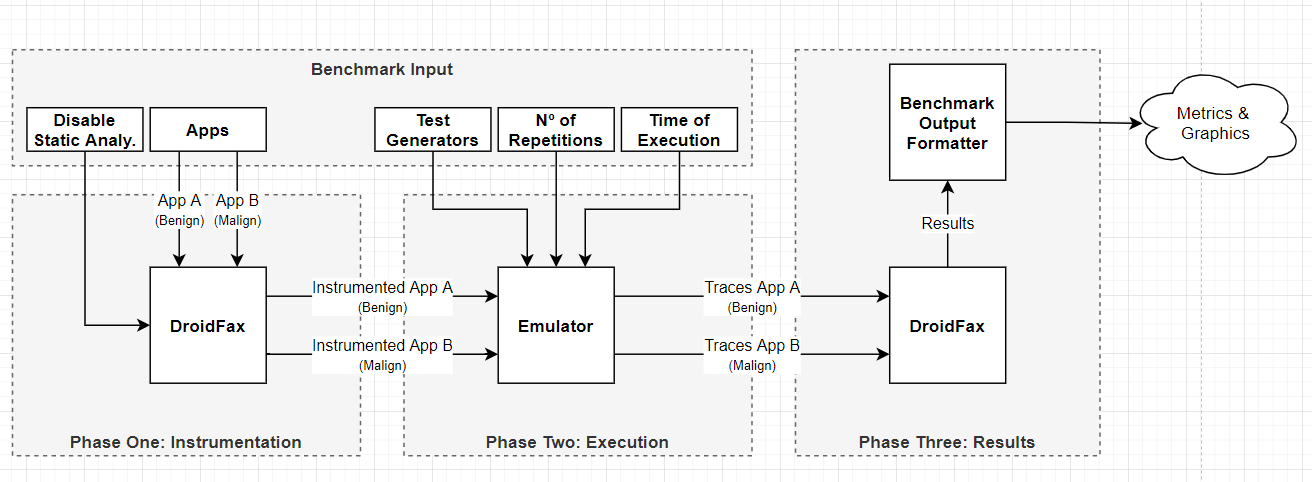}
  \label{benchArq}
  \caption{DroidXP architecture}
  \label{fig:benchArq}
\end{figure*}

Figure~\ref{fig:benchArq} shows the DroidXP architecture, based on the pipes-and-filters
architectural style \cite{architecture-book}. 
The architecture includes three main components; where each component is responsible for a specific phase of the
experiments execution (instrumentation, execution, and result analysis).

\subsubsection{Phase 1: Instrumentation}

In the first phase, a researcher must define the corpus of APK files DroidXP should consider during an
experiment execution. After that, DroidXP starts the DroidFax service that instruments each APK file,
so that DroidXP would be able to collect data (e.g., calls to sensitive APIs) about each execution.
To improve the performance of DroidXP, the instrumentation phase runs only once for each APK.
In this phase, the DroidFax tool also runs some static analysis procedures---when the
option \texttt{--disable-static-analysis} is not set.

\subsubsection{Phase 2: Execution}

In this phase, DroidXP installs an (already instrumented) APK file into
an Android emulator, and then executes a test case generation tool
during a period of time. This process repeats for every test case generation
tool and APK files. To provide repeatability of the experiment, DroidXP
removes all data stored in the emulator before starting
a new execution. That is, every execution uses a \emph{fresh} emulator,
without any information that might have been kept during
previous executions. 
It is relatively easy to add new test
case generation tools into DroidXP. Indeed,
every new tool must override two methods
of a \texttt{Tool} abstract class (according to
the Strategy Design pattern~\cite{patterns-book}.

\subsubsection{Phase 3: Result Analysis}

During the execution of the instrumented apps, all data that is relevant to our
research is collected by Logcat~\cite{Logcat}, one of the Android SDK's native tools. Logcat dumps a log from the Android emulator
while the already instrumented app is in execution. The part of the log we analyze in this phase comprises
the data sent by the methods within the Android app that were instrumented on the first
phase using the DroidFax tool. 

This data includes method coverage from the execution of each test generator tool and the
set of sensitive APIs the app calls during its execution. This set of calls to
sensitive APIs is necessary to estimate the test generator performance in identifying malicious apps---by spotting
differences between the sensitive API accessed by each version of an app (benign or malign).
In the end, DroidXP outputs the results of the experiment, which gives the
performance of one or more testing generator tools in mining sandboxes.

We used the DroidXP infrastructure to conduct our
first empirical study, whose settings we present in the following section. 

\subsection{First Study: A replication of the \blls}\label{sec:set1}

The \blls reports the results of an empirical study that compares the performance of test generation tools to mine Android
sandboxes~\cite{DBLP:conf/wcre/BaoLL18}. Since the \blls does not
compute the possible impact of DroidFax into the performance of the test generation tools,
here we replicate their work to understand the impact of the DroidFax static analysis algorithms into the \blls results.

Our replication differs from the original work in a few decisions. First, here we isolate
the effect of the DroidFax static analysis algorithms, in the task to identify malicious apps. In addition, although we use the same dataset of
$102$ pairs of Android apps used in the \blls, here we discarded $6$ pairs for which
we were not able to instrument---out of the $102$ pairs used in the original work, originally shared in the AndroZoo repository~\cite{DBLP:conf/msr/AllixBKT16}. We also introduced a recent test generator tool (Humanoid ~\cite{DBLP:conf/kbse/LiY0C19}), which
has not been considered in the previous work. Finally, we extended the execution time of each test generation tool,
executing each app from the test generation tool for three minutes (instead of one minute in the
original work),
and built the sandboxes after executing each test generation tool
three times---the original work executed each test generation tool
only once. It is important to note that our goal here is not to conduct an
exact replication of the \blls, but instead understand
the role of the DroidFax static analysis algorithms in the
performance of test case generation tools for mining sandboxes.

Besides Humanoid, our study considers three test generation tools used in the \blls: DroidBot~\cite{DBLP:conf/icse/LiYGC17},
DroidMate~\cite{DBLP:conf/icse/JamrozikZ16}, and Monkey~\cite{Monkey}. We selected DroidBot and DroiMate because they achieved
the best performance on detecting malicious behavior---when considering the $102$ pairs of Android apps (B/M) in the \blls.
It is important to note that here we used a new version of DroidMate (DroidMate-2), since it presents several enhancements
in comparison to the previous version. We also considered the Google's Monkey open source tool, mostly because it is the most
widely used test generation tool for Android~\cite{DBLP:conf/sigsoft/ZengLZXDLYX16}. Monkey is part of the Android SDK
and does not require any additional installation effort. We included Humanoid in our study
because it is a recent tool that emulates realistic users, creating human-like test inputs using deep learning techniques.

\subsubsection{Data Collection}

Similarly to the \blls, besides
method coverage information, our experiments record every call
to \emph{sensitive methods} of the Android
platforms, while a given test case generation tool
is running. We consider the
same set of 97 sensitive methods 
from the AppGuard privacy-control framework
uses~\cite{DBLP:conf/esorics/BackesGHMS13}.

We executed DroidXP using two
configurations. In the first (named WOS), we executed DroidXP
using the dataset of 96 pairs of Android apps---each pair
including a benign and a malign version,
the four test case generation tools (DroidBot, DroidMate, Monkey, and Humanoid),
and the \texttt{--disable-static-analysis} option of
DroidFax, which disables the
execution of the DroidFax static analysis
component from the experiment. The WOS configuration
runs the test case generation tools for three times, using
a time limit of three minutes. 
In the second configuration (named WS), we executed DroidXP
using the same dataset of 96 pairs of Android apps, though 
also executing a fake test case generator tool (named Joker)
{\bf without} the \texttt{--disable-static-analysis} option.
Joker simulates a test tool that does not
run the Android apps during an experiment execution, and its usage
allow us to estimate the actual performance of the DroidFax static analysis component.

Using the WS configuration, the Execution Phase of
DroidXP does not collect any call to 
sensitive APIs, and thus we can estimate the performance of the
static analysis component of DroidFax (answering
RQ1).
Differently, the WOS configuration
disables the static analysis component of DroidFax and
we could better estimate the true performance of the test
case generation tools for mining android sandboxes 
(answering RQ2). For comparison purpose, we also
executed the four test case generation tools using the
DroidFax static analysis component.

\subsubsection{Data Analysis} 

DroidXP produces a
dataset with the sensitive
APIs that the benign / malign
versions of an app call, during
the execution of each test case
generation tool. We estimate the
performance of a test case generation
tool by considering the percentage of
malwares in our dataset
its resulting sandbox is able to identify .

\begin{figure}
  \centering{\includegraphics[scale=0.4]{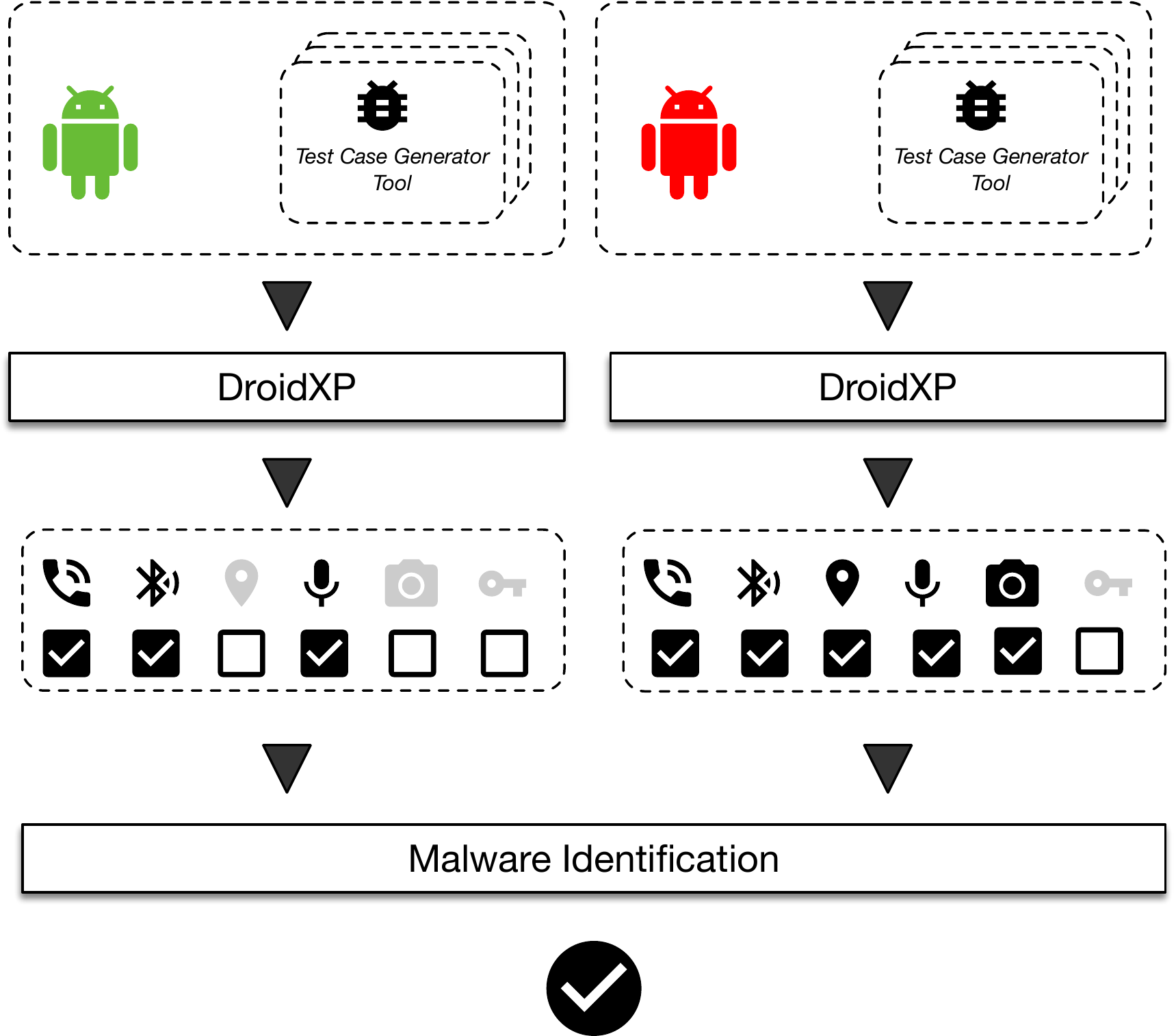}}
  \caption{Overview of our approach for malware identification in the first study.}
  \label{fig:settings1}
\end{figure}

Recall that we build a sandbox 
during the exploratory phase of the mining
sandbox approach. This exploratory
phase records the set of 
sensitive APIs a benign version of an
app calls---during the execution of a test
case generation tool. Similarly to the \blls,
we consider that a sandbox of an
app identifies a malware whenever the
malicious version makes a call to a sensitive API  
that has not been recorded during the exploratory
phase (see Figure~\ref{fig:settings1}).

To sum up, in order to analyse the
performance of the test case generation tools (including Joker),
we just have to compare the calls to sensitive APIs made by
the benign and malign versions of the apps, during
the execution of the tools. In the end, we generate
a set of observations, where each observation
contains the tool name, the number of the repetition (in the
range [1..3]), a boolean value reporting the use of the
DroidFax static analysis component, and a boolean value indicating
whether or not the malware has been identified. We use descriptive statistics
and plots to compare the performance of the tools and
answer RQ1 and RQ2. We also use \emph{Logistic Regression}~\cite[Chapter~4]{statistical-learning}
to understand the statistical relevance and
the contribution of each feature (tool, repetition, DroidFax static analysis
component) to malware identification. Our hypothesis here is that
the DroidFax static analysis component has a positive
effect on the performance of the sandboxes to identify malwares. 

\subsection{Second Study: Use of Taint Analysis for Malware Identification}\label{sec:set2}

In the second empirical study 
we investigate whether or not a taint-based static analysis approach is also promising for
identifying malwares, given a version of an app that we can assume to be secure (goal of research
question RQ3).
To this end, we leverage the FlowDroid
taint analysis algorithms for Android apps (version 2.8), in order to identify dataflows
that might lead to the leakage of sensitive information. Our
goal here is to investigate if it is possible to detect malicious
behavior by means of the \emph{divergent} source-sink paths that FlowDroid reveals after
analysing a benign and a malign versions of an Android app.

\subsubsection{Data Collection}

FlowDroid takes as input an Android Application Package (APK file) and
a set of API methods marked either as {\bf source}
or {\bf sink} (or both). Source methods are those that access \emph{sensitive information} (e.g.,
a method that access the user location), while sink methods are those 
that \emph{might share information with external peers} (e.g., a method that
sends messages to a recipient). We rely on the source-sink definitions
of the FlowDroid implementation~\cite{arzt:pldi-2014,rasthofer-source-sink},
which involves a curate list of source and sink methods (including callbacks and
other Android API methods of interest).
FlowDroid then uses a \emph{context, flow, and field
sensitive analysis} to identify dataflow paths from sources to sinks~\cite{arzt:pldi-2014}.

\begin{figure}
  \centering{\includegraphics[scale=0.4]{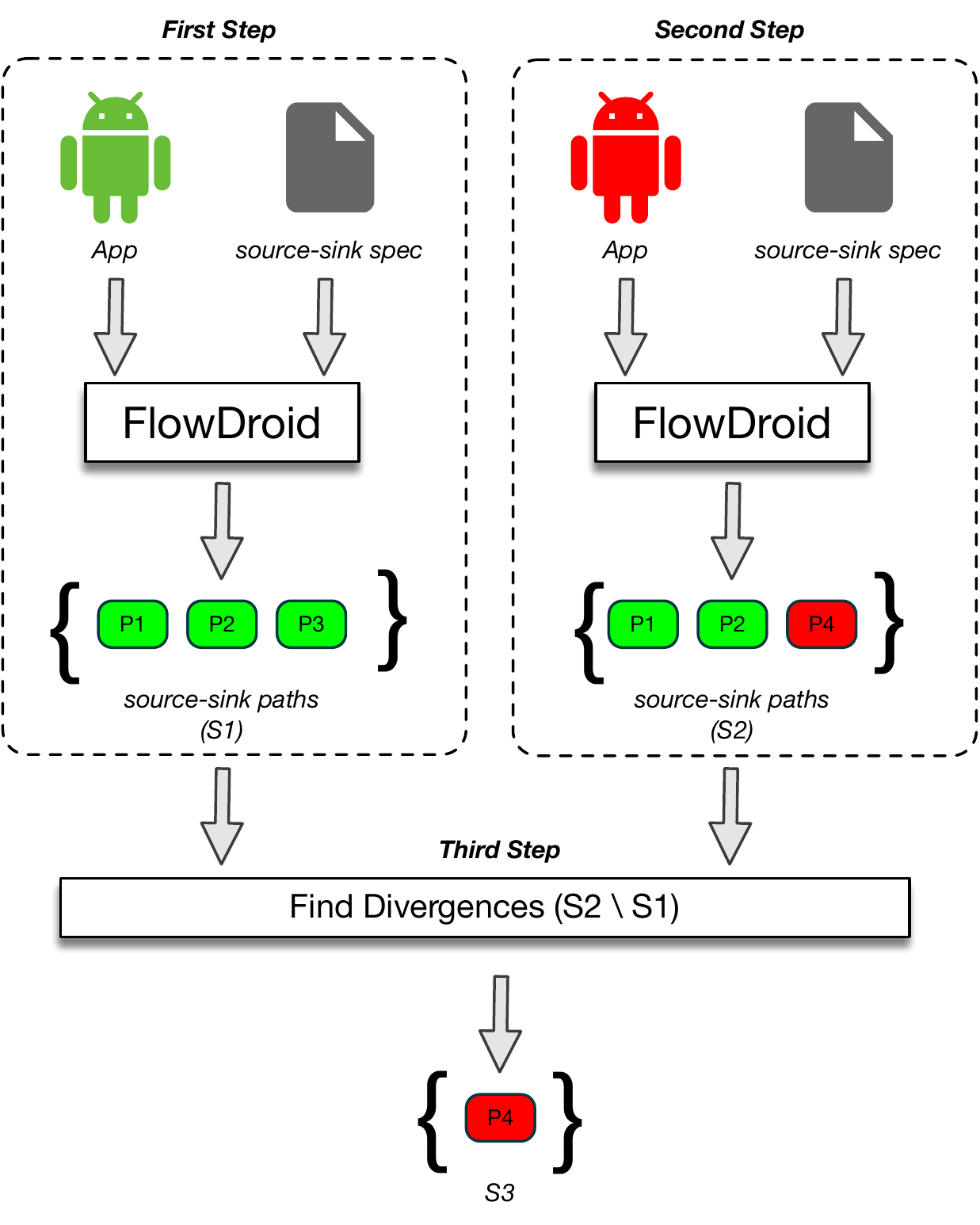}}
  \caption{Overview of our approach in the second study.}
  \label{fig:settings2}
\end{figure}

Our data collection approach involves three steps (see Figure~\ref{fig:settings2}). In the first, we execute FlowDroid to mine the source-sink paths from a benign version of an app, and then enumerate a set (S1) with the 
possible dataflows between sources and sinks. All paths in S1 are considered secure
in our analysis. In the second step we repeat the FlowDroid execution, though
considering the malicious APK version of the app.
This leads to a second set (S2) of source-sink paths.

It is important to note that not all source-sink paths are malign, and then we
follow a specific methodology to identify malwares using taint analysis. That is,
we only report a malware when
FlowDroid finds an additional source-sink path in the malicious version of an app, which
has not been identified when analysing the benign version. Therefore, in the third step we compute the difference (S3) between the sets S2 and S1 (i.e., $S3 = S2 \setminus S1$). If the set S3 is not empty, we assume that FlowDroid
has identified the malware.

In this second study we use the same dataset of $96$ pairs of Android apps (B/M) used in the first empirical
study.

\subsubsection{Data analysis}

We use two metrics in this second study:
the total number of malicious apps FlowDroid is able to
find and the execution time for running the taint analysis algorithm
for each app. Similarly to the first empirical study,
we use descriptive statistics
and plots to compare the performance of the taint analysis and
mining sandbox approaches. We also use \emph{Logistic Regression}~\cite[Chapter~4]{statistical-learning}
to better understand the statistical
significance of the benefits of using FlowDroid
(in comparison to the DroidFax static analysis
component only). Our hypothesis here is that
FlowDroid outperforms, in terms of the number of
detected malware, the sandbox generated
by the DroidFax static analysis component. 

%% file: results.tex
\section{Results and discussion}

In this section we detail the findings of our study. We present the results of the first and
second studies in Section~\ref{sec:res-fs} and Section~\ref{sec:res-ss}, respectively. In Section~\ref{sec:implications} we summarize the
implications of our study. 

\subsection{Result of the first study: A \blls replication}\label{sec:res-fs}

Our first study is a replication of the \blls.
As discussed in the previous section, we first executed the analysis using the DroidXP benchmark with its default
configuration. Then we repeated the process, however this time we isolate the effect of the static analysis component of DroidFax. In this way, we could better estimate the performance of the dynamic analysis tools for mining Android sandboxes.
Table~\ref{tab:fs} summarizes the results of the executions. The columns Exec. (WS) and Exec. (WOS) 
show the number of malwares identified when executing each tool with the
support of the DroidFax static analysis algorithms (WS) and without the support
of DroidFax static analysis algorithms (WOS). 
The Impact column shows 
(in percentage) to what extent the DroidFax static analysis algorithms influences
the performance of the sandboxes created after
executing the test generation tools. We calculate the impact
using Eq. (1).

\begin{eqnarray}
Impact & = & \frac{(Exec.\ (WS) - \ Exec.\ (WOS)) \times 100}{Exec.\ (WS)} 
\end{eqnarray}

Table~\ref{tab:fs} shows that the impact of DroidFax in the results is significant, ranging
from 16.44\% (DroidBot) to 51.79\% (Humanoid). Note that, in the \blls, the authors do not present a
discussion about the influence of DroidFax in the performance of the
test generation tools, even though this influence is not negligible. 
Considering the \joke
tool, our fake test generation tool that does not execute the apps during
the benchmark execution, DroidFax improves the performance in 100\%.
This result is expected, since the \joke tool does not execute any dynamic analysis.
Next we discuss the result of each individual test generation tool. 

\begin{table}[ht]
  \caption{Summary of the results of the first study. }
  \centering
  \begin{small}
 \begin{tabular}{lrrr}
   \toprule
   Tool & Exec. (WS) & Exec. (WOS) & Impact (\%) \\   \midrule
   DroidBot &  73 & 61 & 16.44 \\ 
   Monkey &  71 & 56 & 21.13 \\ 
   DroidMate &  68 & 52 & 23.53 \\ 
   Humanoid &  56 & 27 & 51.79 \\ 
\joke &  42 & 0 & 100.00 \\ 
 \bottomrule
 \end{tabular}
 \end{small}
 \label{tab:fs}
\end{table}

\begin{description}
\item[DroidBot] in the first execution (Exec. WS) led to a sandbox that detected a total of $73$ malware among $96$ pairs present in our dataset ($76.04$\%),
  detecting more apps with malicious behavior than any other tool. Similar to the \blls, DroidBot is the test case generation tool
  whose resulting sandbox detected the largest number of malicious apps. Moreover, in our second execution (Exec. (WOS)), removing the DroidFax
  static analysis support reduced the DroidBot performance in 16.44\%, the smaller impact we observed among the tools.

  \item[Monkey] in the first execution (Exec. (WS)) produced a sandbox that detected $71$ out of the $96$ pairs of Android apps.
    Contrasting, in the original study, the Monkey's sandbox detected $48$ malwares within the 102 pairs ($47.05$\%). This difference
    might be due to the fact that Monkey uses a random strategy for test case generation and here we considered the outcomes
    of three executions---while in the \blls, the authors consider the outcomes of one execution. 
    Considering our second execution (Exec. (WOS)), there is a reduction of $21.13$\% in the Monkey's performance, leading to
    a sandbox that was able to detect $56$ malwares. 

  \item[DroidMate] in the first execution (Exec. (WS)) led to a sandbox that detected 68 apps with malicious behavior ($70.83$\%).
    In the \blls study, DroidMate also detected $68$ malwares, though considering the $102$ pairs of apps used in the
    original study. In the second execution (Exec. (WOS)),
    without the DroidFax static analysis algorithms, the resulting sandbox's performance drops by $23.53$\%, being able to detect
    52 out of the 96 pairs of Android apps.
    
  \item[Humanoid] showed the worst performance, even though a previous work~\cite{DBLP:conf/kbse/LiY0C19} presented that it leads to
    the highest number of lines coverage in comparison to Monkey, DroidBot, and DroidMate. This might suggest that, since Humanoid learn how humans interact with apps, and use the learned model to guide test generation, at simulate environment, this method to generate test inputs are less effective to build Android sandbox, in comparison with techniques that rely on random testing (such as Monkey). In the first execution (Exec. (WS)),
    the resulting Humanoid sandbox identified $56$ malwares in our dataset ($58.33$\%). Humanoid was the most affected in the second
    execution (Exec. (WOS)), whose resulting sandbox presents a reduction of $51.79$\%  in the number of detected malwares.
    Since the \blls did not explore Humanoid,
    we do not have a baseline for comparison with the previous work.

  \item[\joke] is our fake test case generation tool that help us understand the performance of the DroidFax static analysis algorithm for mining sandboxes. 
    We integrated \joke into the DroidXP benchmark as an additional test case generation tool that does not run the Android apps.
    As a result, the analysis using \joke reveals the performance of DroidFax static analysis algorithms alone. For the first execution, with the DroidFax static
    algorithms enabled, even though \joke does not execute the Android apps, its resulting sandbox detected 43.75\% of the malwares. For the second execution,
    that is, disabling the DroidFax static analysis algorithm, the resulting \joke sandbox was not able to detect any malware. Therefore,
    our results show that DroidFax alone is able to detect more than 40\% of the malicious version of the apps. 

\end{description}

\begin{finding}
  Integrating the dynamic analysis tools
  with the DroidFax static analysis algorithms
  improves substantially the performance
  of the resulting Android sandboxes for
  detecting malicious behavior. 
\end{finding}
 
\begin{figure}[htb]
  \centering{
  \includegraphics[trim=60 20 0 50,scale=0.9]{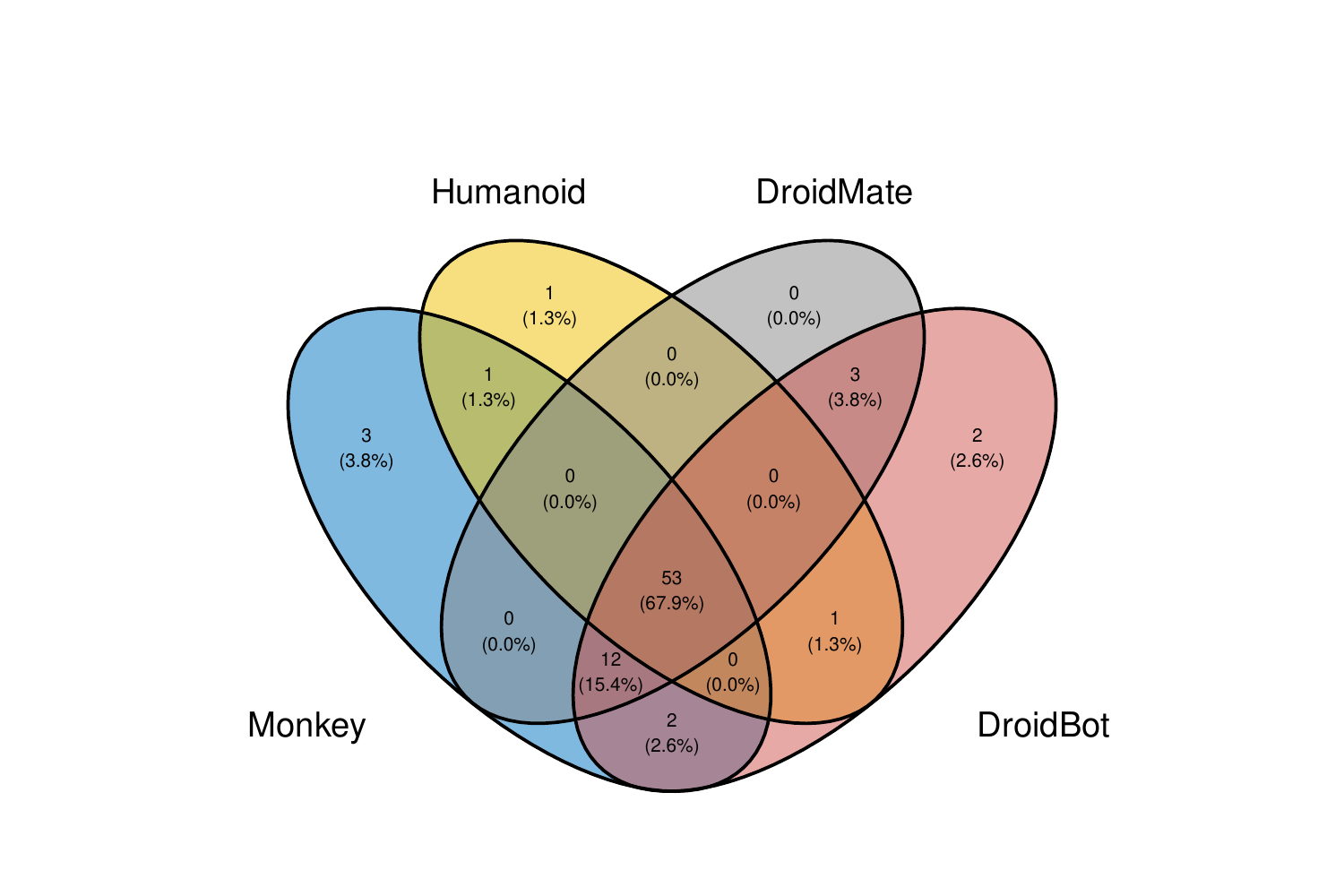}}
  \caption{Venn Diagram highlighting how the sandboxes from the tools can
    complement each other.}
  \label{fig:venn-plot1}
\end{figure}

The Venn-diagram of Figure~\ref{fig:venn-plot1}
summarizes how the tools can complement each other.
Note in the diagram that $53$ malwares have been detected
by all sandboxes generated in the first execution (with the DroidFax static analysis algorithms),
out of the 78 malwares identified by at least one sandbox. In addition, the DroidMate sandbox did not detect
any malware that had not been detected by the other tools. Differently, the Monkey sandbox detected
three malwares that had not been detected by any other sandbox, the DroidBot sandbox detected two malwares
that had not been detected by any other sandbox, and the Humanoid sandbox detected one malware that had not
been detected by any other sandbox. 
Contrasting with the \blls,
our results suggest that using DroidMate in combination with Monkey, DroidBot, and Humanoid
does not improve the general performance of an integrated environment for mining
Android sandboxes.

\begin{finding}
  Our results suggest that one might benefit from using  an integrated
  environment that combines Monkey, DroidMate, and Humanoid to
  mine Android sandboxes. Contrasting with the \blls, introducing the DroidMate 
  tool does not improve the overall performance for detecting malwares using
  a mining sandbox approach.
\end{finding}

Altogether, ignoring  \joke, our study reveals that from $58.33$\% (Humanoid)
to $76.04$\% (DroidBot) of the malicious apps investigated in our study can be
detected using the sandboxes generated after running the test case tools with the support of the
DroidFax static analysis algorithms. We also investigate if the use
  of the DroidFax static analysis component leads to a statistically significant benefit
  on malware identification. To this end, we build a logistic regression model
in the form $Malware \sim Tool + StaticAnalysis + Repetition$. Table~\ref{tab:log-reg1}
shows the results of the logistic regression analysis, highlighting that
(a) Humanoid has a negative, though significant impact on malware identification; and
(b) the use of DroidFax static analysis has a positive and significant
impact on malware identification.

\begin{table}[ht]
  \centering
     \caption{Results of the Logistic Regression (first study)} 
       \begin{small}
     \begin{tabular}{lrlc}
   \toprule
  & \emph{Estimate} & \emph{p-value} & \emph{C.I.} \\ 
   \midrule
   Tool [DroidBot] & 0.1034 & 0.4718 & (-0.133$, $0.340) \\ 
   Tool [DroidMate] & -0.0561 & 0.6955 & (-0.292$, $0.180) \\ 
   Tool [Humanoid] & -0.8910  & 0.0000 $^{***}$ & (-1.131$, $-0.651) \\ 
   Tool [Monkey] & -0.0110 & 0.9390 &  (-0.247$, $0.225)\\ 
   DroidFax static analysis & 0.8867 & 0.0000 $^{***}$ & (0.743$, $1.031)\\ 
   Repetition & -0.0171 & 0.7487 & (-0.105$, $0.071) \\ \midrule
   AIC            & 3001.07     \\
   Num. obs.      & 2304        \\
   \bottomrule
   \multicolumn{4}{l}{\scriptsize{$^{***}$ \emph{p-value} $<0.001$}}
     \end{tabular}
        \end{small}
    \label{tab:log-reg1} 
\end{table}

Besides that, in the first execution (WS), none of the
resulting sandboxes could detect 18 malwares in our dataset ($18.75$\%). According to
the Euphony tool~\cite{hurier2017euphony}, $12$ of these $18$ malwares are \emph{adwares}, $3$ are \emph{trojans}, $2$ are
PUPs (\emph{Potentially Unwanted Program}), and one is an \emph{exploit}.
At this point, an additional question arises: what are the characteristics
of the malwares that have (not) been identified using the mining sandbox approach?
To explore this question, we take advantage of the
\texttt{dex2jar} tool to reverse-engineer all 96 malwares considered in our analysis and
computed the \emph{diffs} of the benign/malicious versions of the APPs.
The results of this activity are available in our replication
package.\footnote{https://github.com/droidxp/paper-replication-package/blob/master/diff/} 
In what follows we dissect a few examples of malwares
that at least one of the resulting  was able to
identify. After that,
we present the characteristics of a malware that none of the sandboxes
was able to detect. Our goal here is to provide
a lower-level intuition about the classes of malware the
mining sandbox approach is not able to detect. A reader that
is not interested in these details could skip to
Section~\ref{sec:res-ss}.

To start with, consider the malicious version of the app \texttt{com.andoop.flyracing}---which both DroidBot and
Humanoid sandboxes could detect in our analysis. 
In this particular case, the malicious version changes the Android Manifest file,
adding permissions to receive and send SMS messages
(Listing~\ref{lst:androidManifest}). Adding these permissions, a malicious app may get money
fraudulently by sending messages without user confirmation, for instance.
The pair \texttt{L:M} indicates
a code segment that appears in line \texttt{L} of the malicious (\texttt{M})
version of an app.

After decompiling this malware, we also observed that the malicious version of the
\texttt{MainService} class introduces a
behavior that collects sensitive information (the International Mobile
Equipment Identity, IMEI) and sends it using an SMS message
(Listing~\ref{lst:mainService}). 

\begin{lstlisting}[caption={Diffs in the \texttt{com.gau.screenguru.finger}
      AndroidManifest file of the malicious
      version}, language=Java,
    basicstyle=\fontsize{8}{6}\selectfont\ttfamily,
    label={lst:androidManifest}]

67:M >    <uses-permission android:name="android.permission.RECEIVE_SMS"/>
68:M >    <uses-permission android:name="android.permission.SEND_SMS"/>
\end{lstlisting}

\begin{lstlisting}[caption={Diffs in the malicious version
      of the class \texttt{com.android.main.MainService}
      (app \texttt{com.gau.screenguru.finger})},
      language=Java, basicstyle=\fontsize{8}{6}\selectfont\ttfamily,
      label={lst:mainService}]

492:M > localObject2 = (TelephonyManager)getSystemService("phone");
493:M > if (localObject2 != null)
494:M > {
495:M >  this.imei = ((TelephonyManager)localObject2).getDeviceId();
496:M >  this.imsi = ((TelephonyManager)localObject2).getSubscriberId();
497:M >  this.iccid = ((TelephonyManager)localObject2).getSimSerialNumber();
498:M > }
// [...]
519:M > if ("".equals(this.destMobile)) {
520:M >  getDestMobile();
521:M > }
522:M > sendSMS(this.destMobile, "imei:" + this.imei)
\end{lstlisting}

The malicious version of the app \texttt{com.happymaau.MathRef} also changes
the Manifest file to require additional permissions as well as change
the behavior of the app (with malicious code). All sandboxes were able to
detect this malware.
In this case, the malicious version of the app changes the Android Manifest file,
requiring permissions to access the network and WiFi states (Listing~\ref{lst:androidManifest2}).
These changes allow an app
to view the status of all networks and make changes to configured WiFi networks.

\begin{lstlisting}[caption={Diffs in the \texttt{com.happymaau.MathRef}
      AndroidManifest file of the malicious
      version}.
      , language=XML,
    basicstyle=\fontsize{8}{6}\selectfont\ttfamily,label={lst:androidManifest2}]

165:M >    <uses-permission android:name="android.permission.ACCESS_NETWORK_STATE"/>
166:M >    <uses-permission android:name="android.permission.ACCESS_WIFI_STATE"/>
\end{lstlisting}

The malicious version also introduces a method \texttt{a},
that actually collects network and WiFi information, like Mac address and the network state
(see Listing~\ref{lst:d}). This information is then shared using an
HTTP request. 

\begin{lstlisting}[caption={Diffs in the malicious version
      of the class \texttt{com.mn.vymq.b.d}
      (app \texttt{com.happymaau.MathRef})},
      language=Java, basicstyle=\fontsize{8}{6}\selectfont\ttfamily,
      label={lst:d}]

105:M > private String a(Context paramContext)
106:M > {
107:M >	String str = ((TelephonyManager)paramContext.getSystemService("phone")).getDeviceId();
108:M > StringBuilder localStringBuilder = new StringBuilder();
109:M > localStringBuilder.append(str);
110:M > paramContext = (WifiManager)paramContext.getSystemService("wifi");
111:M > if (paramContext == null) {}
112:M >  for (paramContext = null;; paramContext = paramContext.getConnectionInfo())
113:M >  {
114:M >   if (paramContext != null)
115:M >    {
116:M >      paramContext = paramContext.getMacAddress();
117:M >      if (paramContext != null) {
118:M >       localStringBuilder.append(paramContext);
119:M >      }
120:M >    }
121:M >    return a(localStringBuilder.toString());
122:M >  }
123:M > }
\end{lstlisting}

All resulting sandboxes also detected the malicious version of the app \texttt{ru.qixi.android.smartrabbits}.
This particular malware also changes the Android Manifest file,
requesting permission to access the location service (Listing~\ref{lst:androidManifest3}).
This permission allows access to location features, such as the Global Positioning System (GPS) on the phone, if it is enabled. Malicious applications can use these features to determine where the phone owner is, which is a
classic and well-documented privacy threat. 

\begin{lstlisting}[caption={Diffs in the \texttt{com.happymaau.MathRef}
      AndroidManifest file of the malicious
      version}.
      , language=XML,
    basicstyle=\fontsize{8}{6}\selectfont\ttfamily,label={lst:androidManifest3}]

8:M >    <uses-permission android:name="android.permission.ACCESS_COARSE_LOCATION"/>
9:M >    <uses-permission android:name="android.permission.ACCESS_FINE_LOCATION"/>
\end{lstlisting}

In addition, the malicious app clandestinely monitors the geographic location of the user and sink
this information to a web server. Listing~\ref{lst:c} shows how
the method \texttt{c}, from the class named \texttt{q}, collects this sensitive information. 

\begin{lstlisting}[caption={Diffs in the malicious version
      of the class \texttt{net.crazymedia.iad.d.q}
      (app \texttt{ru.qixi.android.smartrabbits})},
      language=Java, basicstyle=\fontsize{8}{6}\selectfont\ttfamily,
      label={lst:c}]

65:M > private Location c(Context paramContext)
66:M > {
67:M > try
68:M >  {
69:M >  if (Arrays.asList(paramContext.getPackageManager().getPackageInfo
              (paramContext.getPackageName(),4096).requestedPermissions).contains
              ("android.permission.ACCESS_FINE_LOCATION"))

70:M >   {
71:M >    paramContext = (LocationManager)paramContext.getSystemService("location");
72:M >    Criteria localCriteria = new Criteria();
73:M >    localCriteria.setAccuracy(1);
74:M >    localCriteria.setAltitudeRequired(false);
75:M >    localCriteria.setBearingRequired(false);
76:M >    localCriteria.setCostAllowed(true);
77:M >    localCriteria.setPowerRequirement(1);
78:M >    paramContext = paramContext.getLastKnownLocation
                    (paramContext.getBestProvider(localCriteria, true));
                    
79:M >    return paramContext;
80:M >   }
81:M >  }
82:M >  catch (PackageManager.NameNotFoundException paramContext)
83:M >  {
84:M >   paramContext.printStackTrace();
85:M >   return null;
86:M >  }
87:M >  catch (Exception paramContext)
88:M >  {
89:M >   paramContext.printStackTrace();
90:M >  }
91:M >  return null;
92:M > }
\end{lstlisting}

This pattern of changing the Android Manifest file and including
new method calls characterizes the classes of malwares for which
the mining sandbox approach excels. 
In a different vein, the malicious version of the app
\texttt{com.andoop.flyracing} is among the apps that none of the sandboxes could
detect. Indeed, the malicious version only changes the Android Manifest file,
modifying the meta-data \texttt{ADMOB\_PUBLISHER\_ID}. The AdMob is a monetizing
service provided by Google, and changing the AdMob \emph{publisher identifier} account redirects
the advertisement's revenue to another destination. Based on
this observation, we envision integrating a different approach that reasons
about modifications to the Android Manifest file and that might complement the mining sandbox
approach into the task for detecting malwares; since the mining
sandbox approach is not able to detect malicious packages that
do not introduce new method calls for sensitive APIs.

\begin{lstlisting}[caption={Diff in the file \texttt{com.andoop.flyracing}
      AndroidManifest file of the malicious version.
      \texttt{B} stands for
      the benign version, while \texttt{M} stands for the malicious version.}, language=XML,
    basicstyle=\fontsize{8}{6}\selectfont\ttfamily,label={lst:app65b}]

1:B < <meta-data android:name="ADMOB_\PUBLISHER_\ID"
                     android:value="a14cf7346295891"/>
---
1:M > <meta-data android:name="ADMOB_\PUBLISHER_\ID"
                     android:value="a14f099bfbf3c61"/>
\end{lstlisting}

\subsection{Results of the second study: Use of Taint Analysis for Malware Identification}\label{sec:res-ss}

In this second study we used a taint analysis approach to mine differences between the
benign and malicious versions of the 96 Android apps in our dataset. To this end we leverage the FlowDroid
tool, which tracks how sensitive information flows through the apps using taint analysis algorithms.
Regarding accuracy, the taint analysis approach detected $58$ out of the $96$ pairs in our dataset ($60,42$\%). That is,
using the taint analysis implementation of FlowDroid alone outperforms the Monkey, DroidMate,
and Humanoid sandboxes computed in the second execution (without the DroidFax static analysis
algorithms). This result shows that static analysis algorithms are promising to complement
the mining sandbox approach.

\begin{finding}
  The performance of FlowDroid to identify malicious behavior
  is equivalent to the performance of the
  mining sandbox approach supported by dynamic analysis only---i.e., without
  the DroidFax static analysis algorithms.
\end{finding}

\begin{figure}
  \centering{
  \includegraphics[trim=60 20 0 50,scale=0.7]{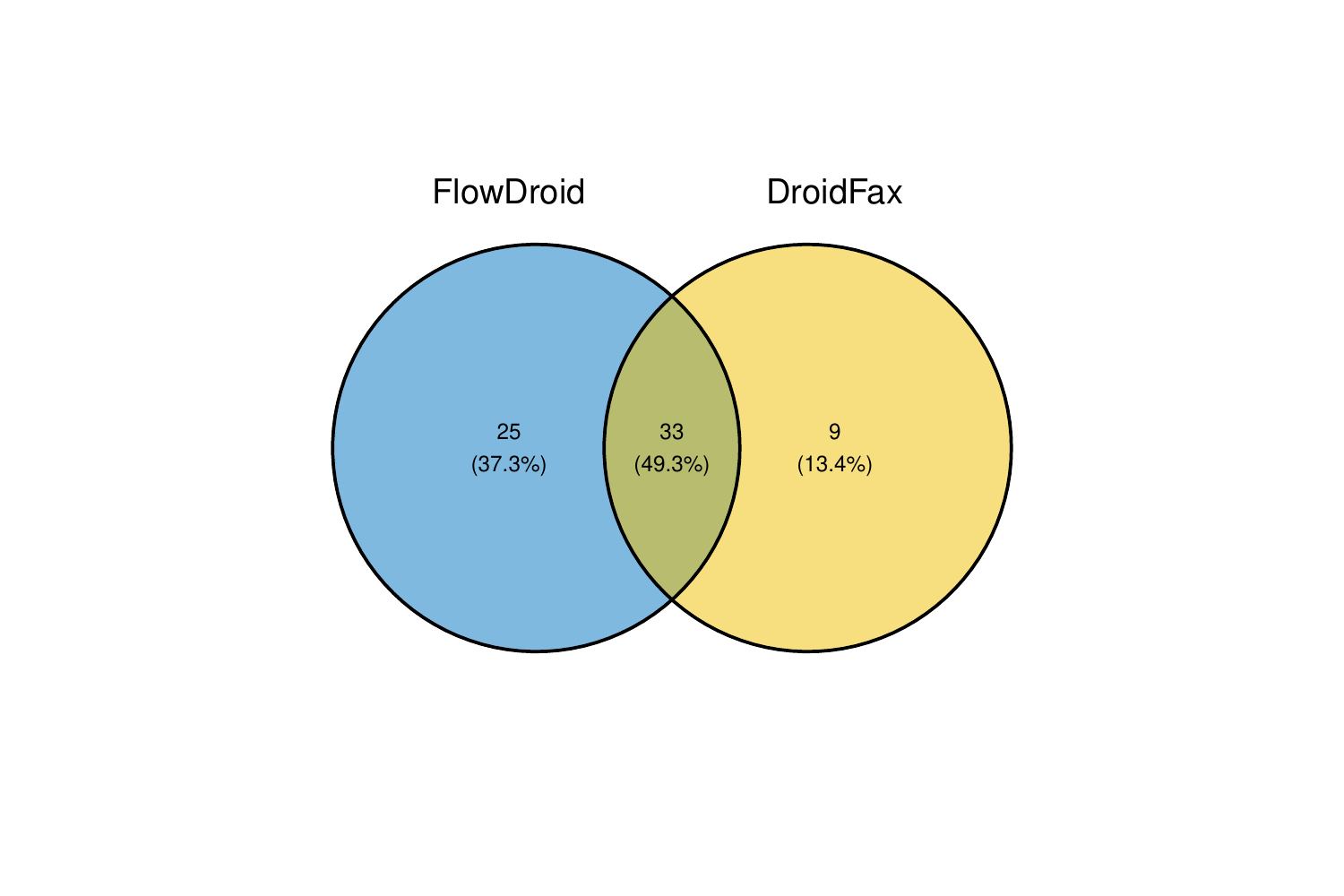}}
  \caption{Venn Diagram highlighting the possible benefits of
    integrating FlowDroid and DroidFax.}
  \label{fig:venn-plot2}

\end{figure}

Additionally, we investigate if we could benefit from combining the
static analysis strategies from FlowDroid and DroidFax. Figure~\ref{fig:venn-plot2} shows a
Venn-diagram summarizing the results. So, when combining
the results from FlowDroid and DroidFax, we were able to detect
$67$ of the malicious apps ($69.79$\%), a result compatible
to the performance we found as response to the first execution of the
test case generation tools---which also considers the DroidFax
static analysis algorithms. More interesting, from those $67$
malicious apps identified, $33$ malwares had been found by
both FlowDroid and DroidFax, even though they follow
a completely different static analysis approach. Furthermore,
FlowDroid shows to be more effective than DroidFax alone, detecting $25$ malicious
apps that had not been detected by DroidFax (while DroidFax detected $9$
malicious apps that had not been detected by FlowDroid). The results
of a logistic regression analysis, considering the
model $Malware \sim Tool$, where Malware is a response
variable indicating if the malware has been detected or not and
Tool is either FlowDroid or the sandbox DroidFax
static analysis component generates, reveals the existence
of a significant difference between
the performance of both tools (see Table~\ref{tab:log-reg2}).

\begin{table}[ht]
  \centering
     \caption{Results of the Logistic Regression (second study)} 
       \begin{small}
     \begin{tabular}{lrlc}
   \toprule
  & \emph{Estimate} & \emph{p-value} & \emph{C.I.} \\ 
   \midrule
   Tool [FlowDroid] & 0.4229 & 0.0428 $^{**}$ & (0.080$, $0.766) \\
   Tool [DroidFax static analysis component] & -1.2730 & 0.0000 $^{***}$ & (-1.560$, $-0.986) \\ \midrule
   AIC            & 334.61     \\
   Num. obs.      & 288        \\
   \bottomrule
   \multicolumn{4}{l}{\scriptsize{$^{***}$ \emph{p-value} $<0.001$; $^{***}$ \emph{p-value} $<0.05$}}
     \end{tabular}
        \end{small}

    \label{tab:log-reg2} 
\end{table}

\begin{finding}
  Integrating the results of static analysis tools
  (such as FlowDroid and DroidFax) seems promising,
  leading to a performance similar to that achieved
  when combining test case generation tools with the
  DroidFax static analysis algorithms. 
\end{finding}

\begin{figure}
  \centering{
  \includegraphics[scale=0.5]{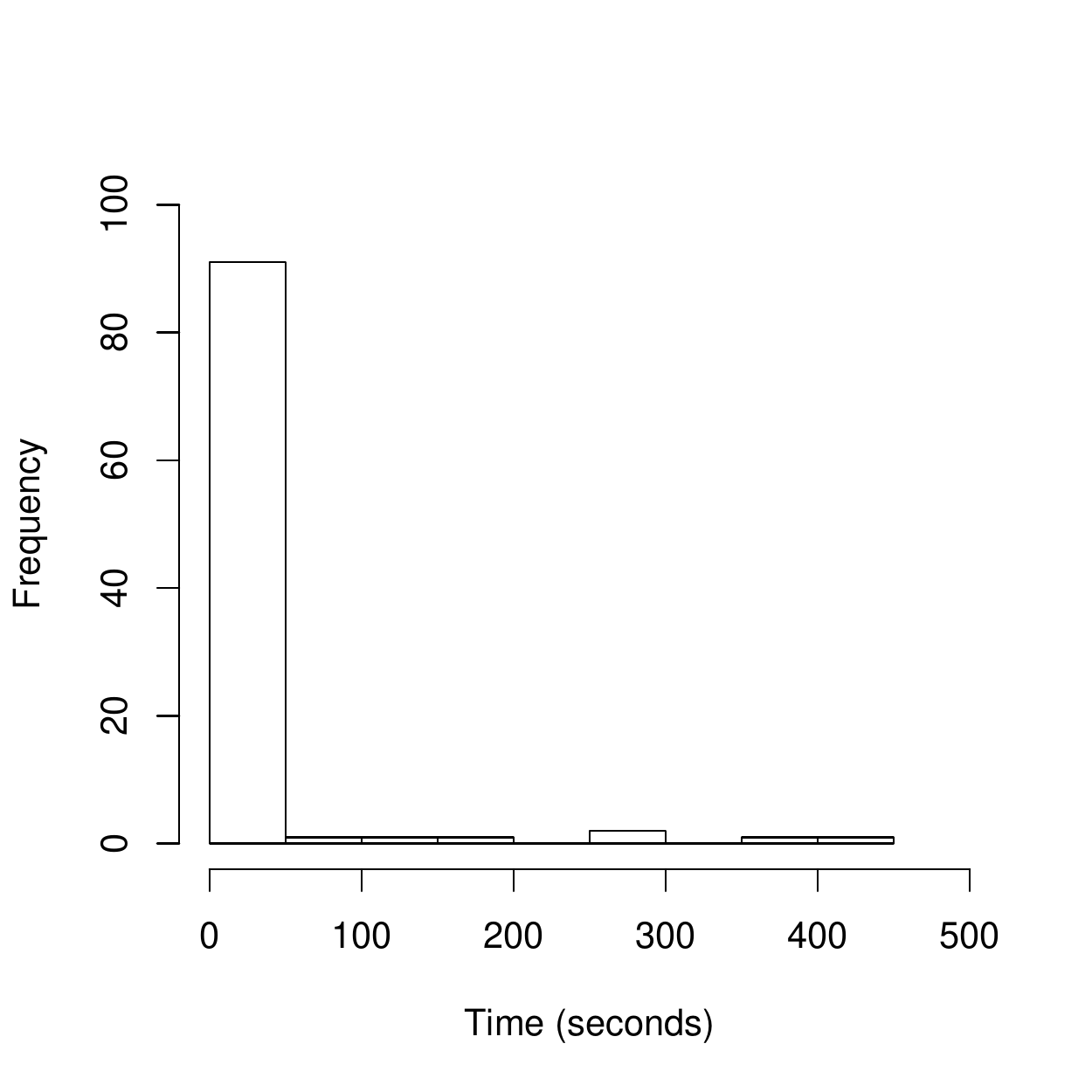}}
  \caption{Histogram summarizing the time to execute FlowDroid}
  \label{fig:histogram}
\end{figure}

The execution of FlowDroid is also feasible: the analysis takes only
32.08 seconds per app on average, totaling a processing time of 52
minutes to analyze all 96 pairs of Android apps.
Even though the time to execute the FlowDroid analysis depends on the size
of the app, the longest run took only 437 seconds. Figure~\ref{fig:histogram}
summarizes the FlowDroid execution time---which most often
concludes the execution in less than 50 seconds (32.11 seconds
on average, with a standard deviation of 70.04). 

Finally, we highlight that FlowDroid was able to detect $4$ malwares among the $18$ malicious Android apps that had not
been detected by the sandboxes constructed in the first study. Among these
four malwares, $2$ are \emph{trojans}, $1$ is an \emph{exploit}, and 1 is an \emph{adware}.

\begin{finding}
  Although FlowDroid presents a performance similar
  to that of using the dynamic analysis approach for mining sandboxes,
  it was able to detect four additional malwares (out of the
  18) that had not been detected in the first study. 
\end{finding}



%% file: implications.tex
\section{Implications}\label{sec:implications}

The results discussed so far bring evidence
that the \blls study overestimated the
performance of the dynamic analysis tools in
malware identification using the mining sandboxes.
This finding has implications for both
researchers and practitioners. First,
we revisit the literature showing that
DroidFax alone is also effective for mining
sandboxes, being able to identify 43.75\%
of the malwares in our dataset. Moreover,
DroidFax identifies malwares that
none of the generated sandboxes
were able to find, increasing the
performance of the sandbox in at most $51.79\%$ (in the
case of Humanoid).

Table~\ref{tab:fs} in the previous section summarizes this finding: when
executing the mining sandbox approach without the
support of DroidFax static analysis, Humanoid's sandbox
could identify only 27 malwares ($28.12\%$ of the
malwares in our dataset).
Conversely, the DroidBot sandbox achieved the best performance in
terms of the number of detected malware without the DroidFax support for static analysis,
being able to identify $63.54\%$ of the malwares.
The message
here is that researchers and practitioners should
explore the use of DroidFax (or a similar tool) in conjunction 
with dynamic analysis techniques for mining sandboxes---
reviewing the findings of the \blls~\cite{DBLP:conf/wcre/BaoLL18}
and enriching the discussion about
the limitations of static analysis for
mining sandboxes~\cite{DBLP:conf/icse/JamrozikSZ16}.

In the second study we used FlowDroid
to explore a novel approach for malware identification,
which aims to compare the source-sink paths of two
versions of an app (one known to
be secure and another that might
have been repackage or that might have
an injected malicious behavior).
Contrasting with the static
analysis limitations discussed in~\cite{DBLP:conf/icse/JamrozikSZ16},
our findings indicate that
this approach is also effective for malware
identification. Indeed, our taint
analysis approach using FlowDroid
detects several malwares
that none of the sandboxes generated
with the dynamic analysis tools (plus
the DroidFax static analysis component)
could identify (see Table~\ref{tab:taint}). These
result has also implications
for both academia and industry. First,
this it reinforces the benefits
of integrating both static and
dynamic analysis for malware identification.
Second, this finding suggests that
practitioners can benefit from using
an integrated approach that
combines the mining sandbox approach
with taint analysis for malware
identification. 

\begin{table}[ht]
\centering
\begin{tabular}{lccc}\toprule
 Test Generation & FlowDroid & Total & \%\\
 Tool & Increase  &  & \\ \midrule
 DroidBot & 6 & 79 & 82.29\\
 Monkey & 7 &  78 & 81.25 \\
 DroidMate & 7 & 75 & 78.12  \\
 Humanoid & 16 & 72 & 75.00 \\
 Joker & 25 & 67 & 69.79  \\\midrule
 
\end{tabular} 
\caption{Malwares detected in 96 pair (B/M) increased by the taint analysis approach}
\label{tab:taint}
\end{table}

%% file: threats.tex
\section{Threats to Validity}

As any empirical work, this work also has
limitations and threats to its validity. We
organize this section using the taxonomy
of Wohlin et al.~\cite[Chapter 8]{10.5555/2349018}.

\textbf{Conclusion Validity} is concerned with
the issues that might compromise the
correct conclusion about the causal relation
between the treatment and the outputs of an
experiment. The use of inadequate statistical
methods and low statistical significance
are examples of threats to the conclusion validity.
Besides using descriptive statistics and
plots, we also leverage
binomial logistic regression to support our conclusions
in our two empirical studies. Indeed, the results of our logistic
regression analysis give evidence about the existence
of a true pattern in the data, indicating that the DroidFax
static analysis component increases the performance of the sandboxes
we built from the execution of the dynamic
analysis tools (first study) and that FlowDroid outperforms
the DroidFax static analysis component in the
task of identifying malwares (second study).

\textbf{Internal Validity} relates to external factors that might
impact the independent variables without the researchers' 
knowledge. Our two empirical studies are \emph{technology-oriented}~\cite{10.5555/2349018,expruna},
which are not subject to learning effect threats. Nonetheless,
due to the random behavior of the test case generation tools,
we should not validate the results of this experiment without considering the presence of
random events in the execution. To mitigate this threat, we have used a configuration of
DroidXP that runs multiple times each tool and computes the average result
from those executions. So, we could adequately compare the results of our experiment with
the results of the \blls. Beyond that,
we tested only 96 of the original 102 pairs of apps in this experiment because the we
could not execute those six pairs of apps due to crashes in the Android emulator.
However, our goal here is not to conduct an exact replication of the previous work,
but actually to better understand how static analysis supports and complements
the mining sandbox approach for malware identification. 

\textbf{Construct Validity} concerns possible issues that might
prevent a researcher to draw a conclusion from the experimental
results. The design of our first study involves one treatment
(a two-level factor indicating the use or not of the
DroidFax static analysis component) and three independent
variables: {\bf app id} (96 level factor), the test
case generation tool (4-level factor, including DroidBot,
DroidMate, Monkey, and Humanoid), and the 3-level factor repetition
(we executed every tool three times for all apps, with and without
the DroidFax static analysis component). The dependent variable
indicates if a malware has been identified by the sandbox
of a given test case generation tool built with (or without) the
DroidFax static analysis component (in a particular repetition). This
design leads to a total of 2304 observations, which is in conformance with the recommendations
of Arcuri and Briand~\cite{arcuri:icse11} for this kind of experiment.
Our second study presents a more straightforward design, comprising a two
factor treatment (FlowDroid x the DroidFax static analysis) and   
the same set of 96 apps of the first study. The dependent variable
indicates if a malware has been identified by FlowDroid
or by the sandbox the DroidFax static analysis component generates. This
design leads to a smaller number of runs,
but we still believe that it is sufficient to draw our conclusions
(as the results of the logistic regression indicate).

\textbf{External Validity} concerns whether or not the
researchers can generalize the results for different scenarios.
Our study shares some of the threats the
\blls had presented. In particular, here we used the same set of pairs of apps from a \emph{piggy-backed} dataset
released by Li et al.~\cite{li2017understanding}. That is, using this dataset, we could not
cover all categories of Android malware. Besides that, we only used a small number of four test
case generation tools in this study. To mitigate these threats and enrich
the generalization of our research, we make available DroidXP, which does allow future
experiments to evaluate other test case generation tools in different malware
datasets.

%% file: conclusion.tex
\section{Conclusions}

In this paper we reported the
results of two empirical studies that explore
techniques for Android malware identification.
The first study is a non-exact replication of a
previous research work~\cite{DBLP:conf/wcre/BaoLL18},
which investigates the Android mining sandbox
approach for malware identification. There,
Bao et al. report that more than 70\% of
the malwares in their dataset can be
detected by the sandboxes built from the
execution of five test case generation
tools (such as Monkey and DroidMate). Our replication
study revealed that this performance is only
achieved if we enable a static
analysis component from DroidFax that was supposed to only
instrument the Android \texttt{apk} files,
though that independently contributes to
building the sandboxes statically. As such,
the use of DroidFax leads to an
overestimation of the performance of the
mining sandbox approach supported by dynamic analysis.
Indeed, the execution of DroidFax alone enabled
us to generate a sandbox that can
identify 43.75\% of the malwares from 
their dataset. 

In the second study we investigated
a new approach based on taint analysis
for malware identification, which
leads to promising results. First,
the taint based static analysis approach
detected 60.42\% of the
malwares in the dataset. When combining
taint analysis with the
mining sandbox approach, we were
able to identify 82.29\% of the
malwares in the dataset. These results
have implications for both researchers and
practitioners. First, we review the literature
showing, for the first time, empirical evidence
that the mining sandbox approach benefits
from using both dynamic and static analysis. Second,
practitioners can improve malware identification using
a combination of 
the mining sandbox approach with taint analysis.
Nonetheless, both the mining sandbox approach and
taint analysis present limitations. In particular,
we are not able to identify a malware that
uses the same set of calls to sensitive APIs of the
benign version of an app, using
the mining sandbox approach. Similarly, we are not
able to identify a malware that presents the same
paths from sources to sinks of the corresponding benign version of
an app, using the taint analysis approach. 
To mitigate these limitations, we envision the use of other approaches---such as machine learning
algorithms to classify changes in non-code assets (e.g., Android
manifest files) and symbolic execution to differentiate
malicious calls or source-sink paths.